\pdfoutput=1
\documentclass[sigconf, screen]{acmart}
\usepackage{pifont}
\usepackage{makecell}

\AtBeginDocument{%
  }

\usepackage{adjustbox}
\usepackage{multirow}

\usepackage{tcolorbox}
\usepackage{listings}
\setcopyright{acmlicensed}
\copyrightyear{2026}
\acmYear{2026}
\setcopyright{cc}
\setcctype{by-nc-nd}
\acmConference[ICCAD '26]{IEEE/ACM International Conference on Computer-Aided Design}{November 08--12, 2026}{San Jose, CA, USA}
\acmBooktitle{IEEE/ACM International Conference on Computer-Aided Design (ICCAD '26), November 08--12, 2026, San Jose, CA, USA}
\acmDOI{10.1145/3831252.3834220}
\acmISBN{979-8-4007-2873-0/2026/11}

\begin{document}


% \title{OpenLayout: Benchmark and Closing the Loop on full automation of VLSI Physical Design}

\title{OpenLayout: An Open and Agentic-Driven Platform for Holistic VLSI Physical Design and Benchmarking}

\author{
  Qiufeng Li$^{1, *}$, Sizhe Tang$^{1}$, Quan Cheng$^{2}$, 
  Zirui Wang$^{3}$, 
  Wantong Li$^{3}$,
  Tian Lan$^{1}$, Weidong Cao$^{1,*}$ \\
  $^{1}$ The George Washington University \quad $^{2}$ Brown University \quad 
  $^{3}$ University of California, Riverside  \\
  Correspondence: {\texttt{\{qiufeng.li, weidong.cao\}@gwu.edu}} \\
  %Agentrys
  % \texttt{quan\_cheng@brown.edu}, \texttt{chengxuanwang@ucla.edu} \\
  % \texttt{dpan@ece.utexas.edu}, \texttt{markren@agentrys.ai}, \texttt{chiatungh@nvidia.com}
}

\renewcommand{\shortauthors}{Li et al.}

\newcommand{\cheng}[1]{{\color{red} #1}}

\begin{abstract}

Physical design automation of very large-scale integrated circuits (VLSI)  has made significant strides, progressing from single-stage placement and routing toward increasingly integrated full-flow implementation. 
Yet, existing approaches struggle with the vast design space optimization (DSO) arising from diverse technology nodes, circuit designs, and varied constraints. 
The substantial runtime of physical-design tools further exacerbates this challenge, leaving industrial workflows heavily dependent on experienced engineers to navigate the design space and achieve power, performance, and area (PPA) targets, incurring significant engineering effort.

We argue that large language models (LLMs) with strong reasoning capabilities, enriched with distilled human design expertise, offer a promising path toward efficient and automated DSO. 
Unlike traditional algorithms, LLMs can interpret complex design contexts and naturally incorporate human expertise expressed in natural language for reasoning, acting, observing, and adapting throughout the design process, a capability that conventional optimization methods cannot exploit. 
To this end, we present \textbf{OpenLayout}, an agentic platform that leverages multi-agent collaboration with embedded design knowledge to enable self-iterative DSO. 
We further introduce \textbf{OpenLayout-Bench}, a benchmark targeting advanced technology nodes with commercial EDA tools to systematically evaluate physical design automation capabilities of LLMs. Experimental results demonstrate that  OpenLayout achieves state-of-the-art performance, with approximately  20\% accuracy improvement over baseline methods. 
Our work is open-sourced at \url{https://github.com/fenggev56/OpenLayout}.

\end{abstract}

\maketitle

\section{Introduction}

\iffalse

Digital VLSI circuits form the foundation of modern electronic systems, powering essential components such as CPUs and GPUs~\cite{vlsi_modern_system, vlsi_modern_systemv2}. 
Growing societal demand for faster and more capable devices places significant pressure on designers to meet stringent PPA targets,  especially on advanced technology nodes under tight schedules~\cite{Fusion_Compiler, innovus}. 
To accelerate time-to-market and improve quality of results, recent works~\cite{Fusion_Compiler, openroad, chateda, dsoai} have pursued full-flow physical design automation, with tools such as Fusion Compiler~\cite{Fusion_Compiler} and OpenROAD~\cite{openroad} as notable examples. 
However, these approaches still struggle with the vast design space optimization (DSO) problem, driven by diverse technology nodes, circuit designs, and varied constraints~\cite{dsoai}.
Further complicating matters, physical design involves deeply interdependent stages,  including floorplanning, macro placement, power delivery network (PDN) synthesis, and clock tree synthesis (CTS), each carrying high runtime costs, making the overall optimization overhead substantial.

\fi

Very large-scale integrated circuits (VLSI) are the foundational hardware cornerstone for advancing nearly all modern critical technologies, such as supercomputing, physical AI, data centers, and quantum computing~\cite{vlsi_modern_system, vlsi_modern_systemv2}. 
The growing demand for more capable hardware places increasing pressure on chip designers to meet stringent power, performance, and area (PPA) targets while adhering to aggressive time-to-market constraints, particularly at advanced technology nodes~\cite{Fusion_Compiler, innovus}. 
To accelerate design cycles and improve quality of results (QoR), recent efforts~\cite{Fusion_Compiler, openroad, chateda, dsoai} have increasingly explored full-flow physical design automation, with industrial and open-source frameworks such as Fusion Compiler~\cite{Fusion_Compiler} and OpenROAD~\cite{openroad} serving as notable examples. 
Despite this progress, these approaches continue to face the formidable design space optimization (DSO) challenge, arising from the diversity of technology nodes, circuit designs, and design constraints~\cite{dsoai}. 
Moreover, physical design is inherently hierarchical and tightly coupled: decisions across floorplanning, macro placement, power delivery network (PDN) synthesis, and clock tree synthesis (CTS) interact in complex ways, while each stage can incur substantial computational cost.
Consequently, efficiently navigating the enormous cross-stage design space while achieving high-quality PPA remains a fundamental challenge for automated physical design.

\begin{figure}[!t]
\centering
% \vskip -15pt
\includegraphics[width=0.75\linewidth]{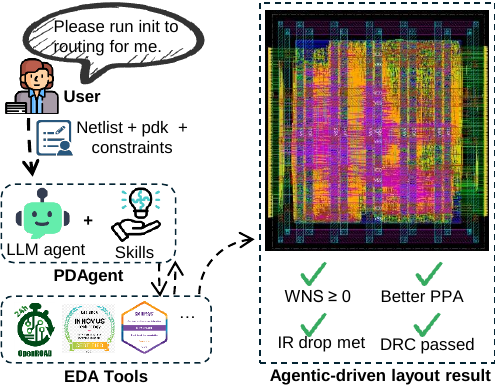}
\vskip -12pt
\caption{OpenLayout takes a netlist, PDK, digital libraries, and constraints as input and autonomously generates DRC-clean, timing-signed-off GDS.
There is no human in the loop. }
\label{fig: overview_1}
\vskip -15pt
\end{figure}

We observe that experienced physical-design engineers navigate this complexity not through exhaustive exploration, but by interpreting design context, analyzing electronic design automation (EDA) tool logs, anticipating how upstream floorplanning decisions affect downstream routing congestion, and adapting strategies to process-specific constraints~\cite{eda_ic, vlsi_modern_systemv2}. 
Such contextual, cross-stage reasoning, typically conveyed through natural language and difficult to encode in conventional optimization algorithms, is largely absent from today’s automated flows. 
We posit that \textbf{agentic AI} offers a natural paradigm for holistic physical-design automation~\cite{llm_eda}, because it can reason, act, observe, and adapt throughout the design process. 
Three structural properties of physical design particularly motivate this view.
\textbf{1) Script-Driven Orchestration.} Each design stage is controlled by parameterized tool commands (e.g., Tcl scripts for commercial place-and-route engines), whose effectiveness depends on expert choices of parameters, pin-assignment strategies, PDN configurations, placement-density targets, and CTS constraints. This script-centric interface provides an actionable mechanism for an agent to translate design intent and domain knowledge into concrete tool actions.
\textbf{2) Structural Regularity.} Designs within the same circuit family (e.g., successive CPU generations) often exhibit recurring architectural and physical-design patterns.
This regularity enables successful strategies and expert knowledge acquired from prior designs to transfer across related instances, reducing the need to optimize each design from scratch.
\textbf{3) Cross-Stage Coupling.} Physical-design decisions are tightly coupled: floorplanning choices influence placement, CTS, and routing, while downstream outcomes can reveal problems that require revisiting upstream decisions.
Effective automation therefore requires a closed-loop process that continuously observes intermediate outcomes and adapts subsequent actions, capabilities that isolated stage-wise optimizers and retrieval-augmented generators alone cannot provide.
A large language model (LLM)-based agentic system can jointly exploit all three properties.
It can translate natural-language domain knowledge into executable design actions, interpret heterogeneous EDA feedback, including logs, timing reports, and congestion metrics, and maintain a coherent decision trajectory across the entire physical-design flow.
This enables a shift from \emph{stage-wise optimization} toward \emph{context-aware, closed-loop automation}, more closely mirroring how experienced engineers reason about and optimize complex physical designs.

\begin{figure*}[!t]
\centering
% \vskip -15pt
\includegraphics[width=0.72\linewidth]{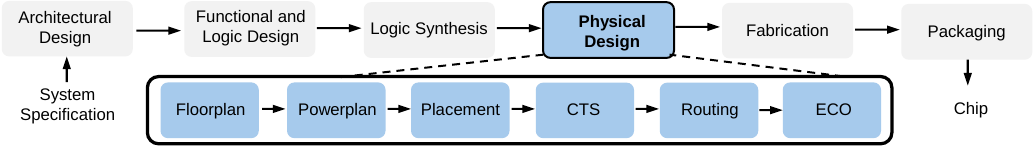}
\vskip -12pt
\caption{Modern VLSI design workflow. The pipeline spans architectural design through packaging, with physical design mainly encompassing floorplan, powerplan, placement, clock tree synthesis (CTS), routing, and ECO.}
\label{fig:workflow}
\vskip -12pt
\end{figure*}

Building on these insights, we introduce \textbf{OpenLayout}, an open-source, multi-agent platform for holistic physical design that formulates layout design as a 
collaborative, self-iterative optimization process driven by reasoning (Fig.~\ref{fig:openlayout_agent}). 
Given a gate-level netlist, technology constraints, digital libraries, and user-specified design objectives, \textbf{OpenLayout} orchestrates a team of specialized agents through a \emph{hub-and-spoke} message-passing architecture to 
produce a DRC-clean, timing-closed layout (Fig.~\ref{fig: overview_1}). 
The framework comprises five purpose-built agents, each encapsulating a distinct concern.
For example, \textbf{Planning Agent} serves as the central hub, decomposing user intent into a directed acyclic graph (DAG) of subtasks, routing all inter-agent messages, and maintaining global execution state.
\textbf{Worker Agent} interfaces with the target EDA tool (e.g., Cadence Innovus) via a persistent interactive shell, executing Tcl commands and returning structured results.
\textbf{Debug Agent} diagnoses failures through LLM-powered root-cause analysis, correlating error signatures against known failure modes.
\textbf{Optimize Agent} performs closed-loop parameter tuning guided by metric trajectories and domain knowledge.
Each agent is equipped with dedicated domain knowledge distilled from the accumulated expertise of senior physical design engineers and researchers.
This knowledge is organized into modular, self-contained \emph{skill
packs}, each encapsulating per-stage expertise: reference Tcl scripts,
parameter tuning strategies, debug decision matrices, and cross-stage
impact analyses. 
Skill packs are automatically discovered during initialization and dynamically retrieved through fuzzy matching, enabling the framework to incorporate new technology nodes and EDA toolchains without modifying the underlying agent logic. 
Together, these components transform physical design from a fixed, script-driven pipeline into an adaptive, knowledge-guided optimization process.

To evaluate our approach, we introduce \textbf{OpenLayout-Bench}, a 
benchmark suite tailored for layout-focused autonomous agents using 
commercial EDA tools at advanced technology nodes, a gap not addressed 
by existing benchmarks~\cite{chateda, iscript}. OpenLayout-Bench consists of two components: 
(1) a holistic flow suite containing 10 circuits of varying 
complexity, 
%including computational logic, communication modules, and large-scale CPU architectures, 
spanning from 
floorplanning to final sign-off; and (2) an atomic task suite covering 
individual implementation steps, such as technology file loading, to evaluate fine-grained agent capabilities.
Our main contributions are summarized as follows:
\begin{itemize}
    \item We introduce \textbf{OpenLayout-Bench}, a benchmark suite for 
    evaluating agent capabilities in physical design using commercial EDA tools at advanced technology nodes. Evaluation results 
    show that standalone LLMs perform poorly on this benchmark; for 
    example, Gemini 3.1-pro-preview~\cite{gemini-3-pro} achieves a pass rate of only 18.9\%, highlighting 
    the difficulty of the task and the need for a dedicated agentic 
    framework.

    \item We present \textbf{OpenLayout}, an open-source, multi-agent 
    platform for holistic physical design automation, covering the 
    complete flow from floorplanning to final sign-off.

    \item Experimental results demonstrate that \textbf{OpenLayout} 
    achieves state-of-the-art performance. Built on Gemini 3.1-pro-preview, our platform achieves a pass rate of 84.4\%, 
    representing a significant improvement over the standalone LLM 
    baseline.
\end{itemize}

\section{Background and Related Work}

\subsection{VLSI Physical Design Flow}

A generic VLSI design flow spans from architectural design to packaging (Fig.~\ref{fig:workflow}). 
Physical design is the most critical and complex design stage.
After logic synthesis converts RTL code to a
gate-level netlist, the physical design stage realizes the netlist as a geometrically placed and routed circuit ready for fabrication, comprising five tightly coupled steps~\cite{eda_ic}.
\textbf{Floorplanning} partitions the die into spatial regions for macros and functional blocks, defines the core area, and determines pin positions along the chip boundary.
\textbf{Power planning} constructs a hierarchical PDN (rings, straps, mesh, vias) to satisfy IR-drop and electromigration constraints with minimal routing overhead.
\textbf{Placement} assigns legal locations to all standard cells and macros, minimizing wirelength and congestion under timing and density constraints.
\textbf{Clock Tree Synthesis (CTS)} builds a buffered clock network that delivers the clock to all sequential elements with bounded skew, latency, and transition time.
\textbf{Routing} assigns wire segments to metal layers for all signal nets, subject to design rule constraints (DRC), while minimizing wirelength, via count, and congestion. 
These stages are strongly coupled: floorplan quality constrains placement, which in turn determines the congestion and timing margins available to CTS and routing.

\subsection{Related Work}

AI has made significant advances in IC design automation ~\cite{rtlcoder, anglogfed, domain-infuse, rose-opt, cktgnn, gao2025analoggenie, wang2025multi, poddar2025heart, gao2025analoggenie-lite,liang2025forgetting}. 
In digital VLSI design, AI-driven methods have been increasingly used for PPA optimization throughout the physical design flow. 
Early approaches can be broadly categorized into two paradigms: \emph{full-flow} optimization, which jointly optimizes multiple stages of the physical-design flow, and \emph{stage-specific} optimization, which targets individual stages such as placement, power planning, CTS, or routing~\cite{ml_eda}. 
We first review these two paradigms, then discuss the emerging use of \textit{LLMs for physical design} and the \textit{benchmarks} used to evaluate existing approaches.

\noindent{\textbf{(i) Full-Flow Optimization.}}
Full-flow optimization seeks to automate multiple physical-design stages jointly. 
Commercial reinforcement learning (RL)-based platforms such as Synopsys DSO.ai~\cite{dsoai} autonomously explore placement, CTS, and routing parameters, while Cadence Cerebrus~\cite{cerebrus} learns transferable AI models for warm starts across designs. 
OpenROAD~\cite{openroad} provides an open-source place-and-route flow on public PDKs.
Yet, these approaches typically rely on predefined parameter spaces and hand-crafted objectives, limiting adaptability to novel designs, constraints, and optimization goals.

\noindent{\textbf{(ii) Stage-Specific Optimization.}}
Substantial effort has also targeted individual physical-design stages. 
\textbf{Floorplanning and Placement.} DREAMPlace~\cite{dreamplace} formulates analytical placement as GPU-accelerated optimization, while AutoDMP~\cite{autodmp} applies Bayesian optimization to macro and standard-cell placement. 
RL-based methods include MaskPlace~\cite{maskplace}, which learns pixel-level placement policies, and Mirhoseini et al.~\cite{graph_place}, which combine GNN-based netlist representations with RL for macro placement. 
\textbf{Power Planning.} OpeNPDN~\cite{openpdn} uses CNN-based PDN template selection with transfer learning, while GridNet~\cite{gridnet} predicts IR-drop and enables localized repair. 
\textbf{CTS and Routing.} GAN-CTS~\cite{gan-cts} combines GANs and RL for CTS parameter optimization, while RouteNet~\cite{routenet} predicts DRC hotspots and congestion to reduce expensive routing iterations. 
Despite their effectiveness, stage-specific methods largely optimize individual stages and do not explicitly capture the strong dependencies between upstream and downstream design decisions.

\noindent{\textbf{(iii) LLMs for Physical Design.}}
The emergence of LLMs introduces a complementary paradigm that shifts from task-specific optimization toward language-driven reasoning and tool interaction. 
Early work applies LLMs to EDA script generation. 
More recently, LLM-based agents have been used to drive complete physical-design flows with open-source EDA tools, including ChatEDA~\cite{chateda} and EDAid~\cite{EDAid}. 
However, existing approaches primarily rely on LLM parametric knowledge, optionally augmented with retrieval-augmented generation (RAG), to interpret tool feedback and generate commands. 
As shown by our experiments (Section~\ref{rag_baseline}), this paradigm struggles with complex designs and advanced technology nodes, where the design space is large and design rules, tool behavior, and closure strategies are highly process-specific. 
These limitations motivate systematic evaluation of whether LLM agents can effectively acquire and apply such domain knowledge across the full design flow.

\noindent{\textbf{(iv) Benchmarks.}}
Existing EDA benchmarks are predominantly stage-specific
For example, the GSRC and MCNC suites~\cite{kahng_vlsi} remain widely used for floorplanning and macro placement.
Yet, they lack the scale and complexity of modern SoCs.
At the flow level, OpenROAD Flow Scripts (ORFS)~\cite{openroad} provides a reproducible RTL-to-GDSII framework using open-source designs and public PDKs such as NanGate45, and has become a common baseline for open-source full-flow PPA evaluation. 
However, the provided benchmarks largely focus on open-source tools and mature technology nodes, leaving advanced-node designs with commercial EDA tools, where design rules and constraints are substantially more stringent, without a standardized evaluation framework. 
This gap motivates \textbf{OpenLayout-Bench}, which targets advanced technology nodes and commercial EDA tools to enable systematic evaluation of agentic physical-design automation.

\section{Benchmark Suite}

\iffalse

We propose \textbf{OpenLayout-Bench}, a benchmark suite designed to evaluate LLM agent capabilities in performing physical design on advanced technology nodes on a variety of VLSI designs. OpenLayout-Bench comprises two complementary components: the 
\textbf{Atomic Task Suite}, which evaluates step-level command 
proficiency across individual physical design stages, and the 
\textbf{Holistic Flow Suite}, which requires agents to execute 
the complete RTL-to-GDS flow end-to-end on diverse VLSI designs.
Atomic Task Suite is distilled from tape-out-validated scripts whose results have been published at top-tier conferences including ISSCC~\cite{isscc}, VLSI Symposium~\cite{vlsi}, A-SSCC~\cite{asscc}, and CICC~\cite{cicc}.
The full-flow suite provides netlist files and requires agents to execute the complete physical design flow end-to-end.

\fi

We propose \textbf{OpenLayout-Bench}, a benchmark suite for systematically evaluating LLM agents on physical design across advanced technology nodes and diverse VLSI designs. 
It comprises two complementary components: the \textbf{Atomic Task Suite}, which evaluates step-level command proficiency across individual physical-design stages, and the \textbf{Holistic Flow Suite}, which evaluates an agent's ability to execute the complete RTL-to-GDS flow end-to-end.
The Atomic Task Suite is distilled from tapeout-validated design scripts whose results have been published at leading conferences, including ISSCC~\cite{isscc}, VLSI Symposium~\cite{vlsi}, A-SSCC~\cite{asscc}, and CICC~\cite{cicc}.
The Holistic Flow Suite provides gate-level netlists and associated design inputs, requiring agents to autonomously navigate the full physical-design flow across diverse designs and technology nodes.

\subsection{Atomic Task Suite}

\begin{figure}[!t]
\centering
% \vskip -15pt
\includegraphics[width=0.70\linewidth]{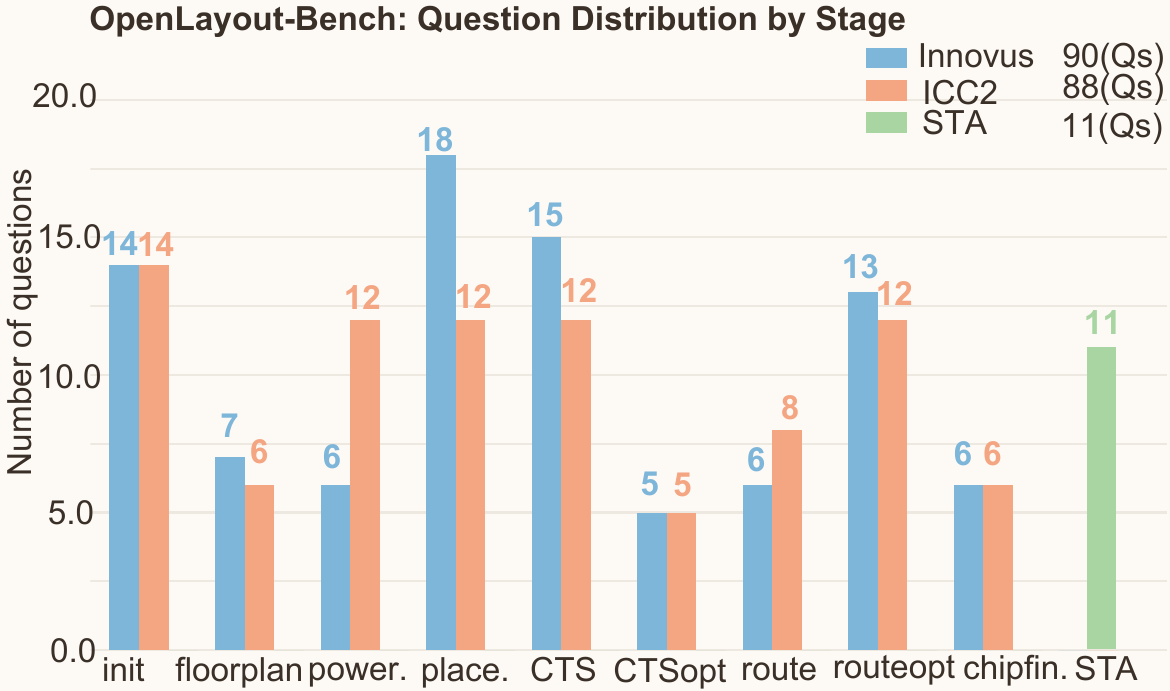}
\vskip -12pt
\caption{Distribution of benchmark questions across physical implementation stages for the Cadence Innovus and Synopsys ICC2 commercial flows. Static timing analysis intentionally over-sampled beyond its stage count to reflect its critical role in industrial sign-off closure.}
\label{fig:atomic}
\vskip -15pt
\end{figure}

The Atomic Task Suite evaluates agent proficiency at the individual 
command and script level, covering the canonical steps of a complete physical design flow. 
Task categories include technology 
LEF/LIB ingestion, MMMC timing constraint setup, core area and utilization specification, power grid definition, and sign-off metric extraction.
One example is shown in Fig.~\ref{fig:example_task}. 
The distribution of questions is weighted by stage 
complexity and industrial importance, with denser coverage in static timing analysis (STA) and physical verification (PV) to reflect their disproportionate cost and criticality in industrial 
sign-off flows. 
The task distribution across physical design stages is shown in Fig.~\ref{fig:atomic}. 
The STA task subset is curated from STA-focused questions used in physical design engineer interviews. 
To avoid data contamination, we retain only questions without a single deterministic answer. 
To ensure correctness, all reference answers are independently validated by three senior physical design engineers, whose consensus serves as the ground truth for evaluation.
The Atomic Task Suite targets commercial flows (Cadence Innovus and Synopsys ICC2) and commercial process nodes (e.g., TSMC 28\,nm), enabling 
evaluation under industrial-grade constraints that more faithfully reflect real deployment conditions.

\begin{figure}[!t]
\centering
\small
\begin{tcolorbox}[title=Example Task: OCV Derate Configuration,
                  colback=blue!5, colframe=blue!40, coltitle=black, fonttitle=\bfseries]

\textbf{Context:} To account for On-Chip Variation (OCV) during the pre-CTS stage, timing derates must be applied to introduce pessimism. For setup checks, the launch path (clock and data) is made \textit{late} (slower) and the capture path is made \textit{early} (faster). Because hold optimization typically occurs after CTS, both setup and hold views are initially set to the \texttt{cmax} corner.

\vspace{0.3em}
\textbf{Question:} Provide the Innovus commands to set the analysis view for both setup and hold to \texttt{core\_view\_cmax}. Additionally, for the \texttt{core\_corner\_cmax} corner, apply a clock derate of 0.952 (early) and 1.048 (late), and a data derate of 1.081 (late).

\vspace{0.3em}
\textbf{Answer:}
\begin{lstlisting}[basicstyle=\ttfamily\small, breaklines=true, backgroundcolor=\color{blue!5}]
set_analysis_view -setup {core_view_cmax} -hold {core_view_cmax}
set_timing_derate -delay_corner {core_corner_cmax} -early 0.952 -late 1.048 -clock
set_timing_derate -delay_corner {core_corner_cmax} -late 1.081 -data
\end{lstlisting}

\vspace{0.3em}
\textbf{Verification:} Exact Match

\end{tcolorbox}
\vskip -15pt
\caption{An example task from the Timing Analysis category (OCV Derate Configuration).}
\label{fig:example_task}
\vskip -15pt
\end{figure}

\subsection{Holistic Flow Suite}

The Holistic Flow Suite comprises 10 open-source RTL designs spanning varying design complexity, organized into three categories: Processor Cores, Arithmetic \& Accelerators, and 
Control \& Communication Circuits, as summarized in 
Table~\ref{tab:benchmark_designs}.
\textbf{Processor Cores.}
TinyRISCV~\cite{tinyriscv} is a lightweight 32-bit RV32IM 3-stage in-order core for microcontroller workloads; Open910~\cite{910} is a high-performance
out-of-order RV64GC core representative of large, macro-rich designs with deep timing paths.
\textbf{Arithmetic \& Accelerator Kernels.}
A parameterized systolic array~\cite{systolic_array_rtl} and a 16b pipelined multiplier cover regular compute-dense circuits.
NVDLA~\cite{nvdla} is included in both small and large configurations, spanning medium- to large-scale accelerator complexity.
AES-256~\cite{aes} provides a cryptographic circuit with structured regular logic and strict
timing constraints.
\textbf{Control \& Communication.}
A UART controller~\cite{uart}, an Ethernet MAC~\cite{ethmac_opencores}, and an elevator finite state machine (FSM) represent I/O-intensive and control-dominated designs, each posing distinct challenges in pin placement and state encoding.

% Add to preamble:
% \usepackage{tcolorbox}
% \usepackage{listings}

% Add to preamble:
% \usepackage{tcolorbox}
% \usepackage{listings}

\subsection{Evaluation Metrics}

\textbf{Implementation Task Suite.}
For individual implementation tasks, we adopt \textit{pass@1} and \textit{pass@5} accuracy as primary metrics, measuring whether the agent produces a correct, executable solution on the first attempt and within
five attempts, respectively.
\textbf{End-to-End Layout Implementation.}
For full RTL-to-GDS flow evaluation, we record two complementary dimensions per design. 
First, \emph{stage completion status}: a binary
pass/fail indicator for each flow stage (init, floorplan, placement, CTS, routing, and sign-off), capturing whether the agent successfully
advances the flow to completion. 
Second, \emph{layout quality metrics}
at flow completion: total cell area ($\mu m^2$), core utilization (\%), worst negative slack (WNS) for timing, and design rule violation (DRV) count. 
Together, these metrics distinguish an agent that
merely executes a flow from one that produces a PPA-optimized, sign-off-clean layout.

\begin{table}[]
\caption{Benchmark designs used for evaluation, categorized by design types with corresponding cell counts.}
\small
\label{tab:benchmark_designs}
\vskip -9pt
\begin{tabular}{lll}
\hline
Category                                   & Design               & \#Cells   \\ \hline
\multirow{2}{*}{Processor Cores}           & TinyRISCV            & 15012     \\
                                           & Open910              & 754,981   \\ \hline
\multirow{5}{*}{Arithmetic \& Accelerator} & Systolic array       & 41198     \\
                                           & 16b pipelined mult.
                                
                                & 342      \\
                                           & AES-256              & 15516     \\
                                           & NVDLA-small          & 270,072   \\
                                           & NVDLA-large          & 1,478,865 \\ \hline
\multirow{3}{*}{Control \& Communication}  & UART controller      & 368       \\
                                           & Elevator FSM         & 176       \\
                                           & Ethernet MAC             & 35172     \\ \hline
\end{tabular}
\vskip -12pt
\end{table}

\section{OpenLayout Framework}
\label{sec:framework}

\begin{figure*}[!t]
\centering
% \vskip -15pt
\includegraphics[width=1.0\linewidth]{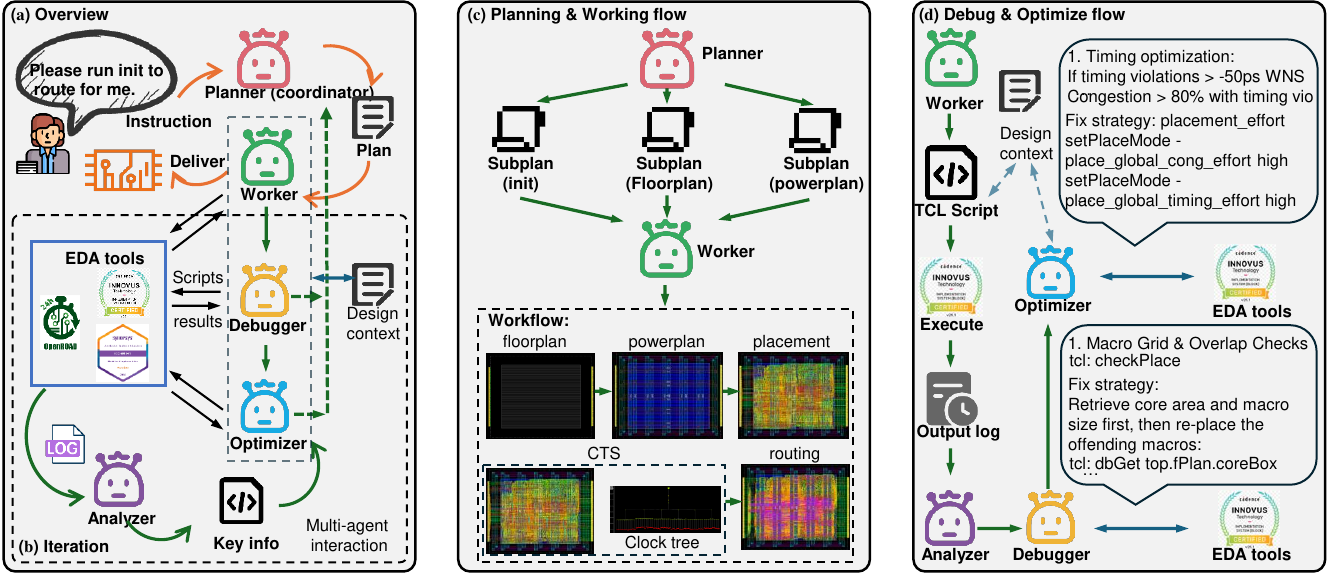}
\vskip -12pt
    \caption{
        \textbf{Overview of the OpenLayout Platform.}
        \textbf{(a)} OpenLayout employs a hierarchical multi-agent system comprising a \textit{Planner} (coordinator), \textit{Worker}, \textit{Debugger}, and \textit{Optimizer}, augmented by \textit{Analyzer}, where the Planner decomposes user instructions into structured plans. 
        \textbf{(b)} Agents iteratively interact with EDA tools (e.g., Innovus) by submitting Tcl scripts, collecting execution results, and extracting key information via the \textit{Analyzer} for closed-loop refinement.
        \textbf{(c)} The planning \& working flow: The Planner decomposes tasks into subplans (init, floorplan, powerplan, etc.), which the Worker executes sequentially through the full physical design flow.
        \textbf{(d)} The debug and optimize flow: the \textit{Debugger} performs checks (e.g., macro overlap) and applies fix strategies, while the \textit{Optimizer} resolves timing and congestion optimization, both guided by the Design Context containing accumulated design information and previously validated scripts.
    }
\label{fig:openlayout_agent}
  \vskip -6pt
\end{figure*}

\subsection{Overview}
\label{sec:overview}

\iffalse

OpenLayout is a multi-agent framework for autonomous physical design that formulates the layout generation problem as a collaborative, iterative optimization process based on reasoning, illustrated in Fig.~\ref{fig:openlayout_agent}(a) and(b).
Given a gate-level netlist, technology constraints, digital libraries, and design objectives, the system orchestrates five specialized agents through a \emph{hub-and-spoke} message-passing architecture to produce a DRC-clean, timing-closed layout.
Following the natural structure of the physical design workflow, we decompose the process into four major phases: planning, implementation, debugging, and optimization. 
These phases form a closed loop that iteratively refines the design until convergence.
Accordingly, we construct a multi-agent system aligned with this structure, comprising 
a Planning Agent, Worker Agent, Debugger Agent, and Optimizer Agent, augmented by an Analyzer Agent that extracts key metrics from verbose EDA tool logs.

\fi

\textbf{OpenLayout} is a multi-agent framework for autonomous physical design that formulates layout design as a collaborative, iterative optimization process driven by reasoning, as illustrated in Figs.~\ref{fig:openlayout_agent}(a) and (b). 
Given a gate-level netlist, technology constraints, digital libraries, and design objectives, \textbf{OpenLayout} orchestrates five specialized agents through a \emph{hub-and-spoke} message-passing architecture to generate a DRC-clean, timing-closed layout.

Following the natural structure of physical design, we organize the process into four phases: \emph{planning, implementation, debugging,} and \emph{optimization}. These phases form a closed loop, enabling the system to iteratively diagnose failures, refine design decisions, and drive the layout toward convergence. Accordingly, OpenLayout comprises a \textbf{Planning Agent}, \textbf{Worker Agent}, \textbf{Debugger Agent}, and \textbf{Optimizer Agent}, augmented by an \textbf{Analyzer Agent} that extracts actionable metrics from verbose EDA tool logs. Together, these agents provide coordinated planning, execution, diagnosis, analysis, and optimization across the full physical-design flow.

\noindent\textbf{Agent Roles.}
Each agent encapsulates a distinct concern:
\textbf{(i)}~\emph{Planner} serves as the central hub, decomposing user intent into a directed acyclic graph (DAG) of subtasks, routing all inter-agent messages, and maintaining global execution state;
\textbf{(ii)}~\emph{Worker} interfaces with the EDA tool (e.g., Innovus) via a persistent interactive shell, executing TCL commands and returning structured results (Fig.~\ref{fig:openlayout_agent}(c));
\textbf{(iii)}~\emph{Analyzer} extracts quantitative metrics, such as timing (WNS/TNS), physical (DRC, utilization), and power, from EDA logs;
\textbf{(iv)}~\emph{Debugger} diagnoses failures through LLM-powered root-cause analysis;
\textbf{(v)}~\emph{Optimizer} performs parameter tuning guided by metric trajectories and domain knowledge, as shown in Fig.~\ref{fig:openlayout_agent}(d).
\noindent\textbf{Hub-and-Spoke Communication.}
All inter-agent messages are routed through the Planning Agent, which 
maintains full observability of the system state and enforces execution 
ordering. The sole exception is \emph{tool interaction}: any agent may 
directly invoke EDA tool commands via a synchronous 
\texttt{execute\_tool} interface, bypassing hub routing to avoid 
latency overhead for frequent tool queries. This hybrid topology 
balances centralized coordination with efficient tool interaction.
\noindent\textbf{Skill Packs.}
Domain knowledge is organized into \emph{skill packs}, self-contained directories structured as \texttt{skills\_\{tech\}\_\{tool\}/} (e.g., \texttt{skills\_tsmc28nm\_innovus/}).
Each skill pack contains per-stage skill files encoding expert knowledge: reference TCL scripts, parameter tuning strategies, debug decision matrices, and cross-stage impact analyses.
Skills are auto-discovered at initialization and resolved dynamically via fuzzy matching, enabling seamless extension to new technology nodes and EDA tools without modifying agent code.
\noindent\textbf{Privacy-Preserving Deployment.}
OpenLayout never exposes proprietary design data to LLMs: technology libraries, netlists, and raw EDA logs remain local, and only distilled metrics (e.g., PPA, DRC counts) are sent to the model.
Agents thus act on the design solely through local TCL execution (see~\cite{pdagent} for more details).
% All sensitive artifacts remain inside the user's environment see~\cite{pdagent} for more details.

% \textcolor{red}{WD: cite your PDAgent-bench work to mention that the framework works in a privacy-preserving manner.}

\subsection{Agent Design}
\label{sec:agent_design}

Each agent inherits from a common \texttt{BaseAgent} class that 
provides message handling, LLM access, tool query primitives, and a 
shared \texttt{DesignContext}. The \texttt{DesignContext} is a cached 
representation of the active design, populated by querying the EDA 
tool after initialization, including macro locations, die/core area, 
and clock topology. It is shared across all agents to maintain 
consistent design state throughout the optimization loop.

\noindent\textbf{Worker Agent.}
The Worker Agent maintains a persistent EDA session and executes 
subtasks in three modes: (1) skill-guided generation, where the LLM 
synthesizes Tcl scripts from skill content and design context; 
(2) script execution, where reference Tcl scripts are sourced and run 
directly; and (3) optimization application, where targeted 
optimizations requested by the Optimizer Agent are applied and 
coordinated through the Planner Agent. Executed scripts, logs, and 
layout snapshots are organized by stage in a session history with 
manifest tracking, ensuring full traceability and enabling downstream 
agents to reference artifacts from any prior stage.

\noindent\textbf{Analyzer Agent.}
The Analyzer Agent is selectively activated at analysis-critical stages 
(e.g., placement, CTS, and routing), where raw EDA tool logs are 
verbose and difficult to interpret directly. At each activated stage, 
it extracts a structured \texttt{StageMetrics} vector covering timing, 
utilization, routing, clock, and power metrics, and passes this 
condensed summary to the Debugger and Optimizer Agents for closed-loop 
refinement.
Lightweight stages such as floorplanning and powerplanning bypass analysis, reducing runtime overhead without sacrificing QoR at critical design junctures.

\noindent\textbf{Debugger Agent.}
After the Worker Agent completes a stage, the Debugger Agent checks 
for DRC violations and other errors by executing an iterative 
diagnosis loop. It loads the stage-specific debug skill, initializes 
a multi-turn LLM conversation with the skill content, EDA output, and 
convergence state as context, then iteratively performs 
check$\rightarrow$fix$\rightarrow$check cycles via direct tool queries. Each step is 
tracked in an \texttt{AgentStageContext} that maintains the full 
conversation history, violations found and fixed, and budget 
consumption. Skills provide the check criteria and corresponding fix 
strategies for each stage.
To prevent unbounded iteration, a three-tier budget hierarchy is 
enforced: (1) micro-level, handling syntax errors with up to 2 
retries; (2) meso-level, handling parameter adjustments with up to 
4 retries; and (3) macro-level, handling cross-stage rollback with 
up to 2 retries. When the budget is exhausted, the Debugger Agent 
escalates to the Optimizer Agent for more aggressive intervention.

\noindent\textbf{Optimizer Agent.}
The Optimizer Agent operates in two modes. In iterative mode, it 
conducts a multi-turn LLM-guided loop analogous to the Debugger Agent: 
checking current metrics, adjusting parameters (e.g., placement 
effort illustrated in Fig.~\ref{fig:openlayout_agent}(d)), informed by stage-specific optimization 
skills and the full iteration trajectory stored in \texttt{MetricsDB}, 
then verifying improvements. In tree-of-thought mode, triggered on 
escalation or LLM-directed exploration, it performs branching design 
space exploration by generating multiple solution candidates and 
executing them either in parallel or sequentially, as described in 
Section~\ref{sec:dso}.

\subsection{Closed-Loop DSO Mechanism}
\label{sec:dso}

The DSO mechanism in OpenLayout operates at two levels: a 
\textbf{stage-level} that progresses through the physical design flow, and a \textbf{parameter-level} that explores 
the design space within each stage.

\noindent\textbf{Stage-Level Execution.}
The Planning Agent maintains a plan $\mathcal{P} = \{s_1, \ldots, 
s_n\}$ of subtasks ordered by dependency. For each subtask $s_i$, 
execution follows one of two paths based on the tool output:

\noindent(1) \textbf{Success path:} Tool $\xrightarrow{\text{result}}$ Planner 
$\xrightarrow{\text{if worthy}}$ Analyzer $\xrightarrow{\text{metrics}}$ 
Optimizer $\xrightarrow{\text{decision}}$ Planner $\rightarrow$ 
advance or re-run;

\noindent(2) \textbf{Failure path:} Tool $\xrightarrow{\text{error}}$ Planner 
$\rightarrow$ Debugger $\xrightarrow{\text{action}}$ Planner 
$\rightarrow$ \{retry, fix, rollback, skip, escalate\}.
The convergence state, classified as improving, stalled, or diverging 
based on the metric trajectory across iterations, informs action 
selection at each decision point.

\noindent\textbf{Parameter-Level Exploration.}
When the Optimizer Agent determines that local parameter adjustments 
are insufficient, for example upon escalation from the Debugger Agent 
or detection of a stalled convergence state, it activates a 
tree-of-thought search over the parameter space. The search proceeds 
in batches: at each batch $b \in \{1, \ldots, B\}$, the LLM generates 
$K$ candidate parameter configurations $\{c_1^{(b)}, \ldots, 
c_K^{(b)}\}$, such as redoing macro placement or revising the power 
plan to reduce routing layer pressure, each accompanied by the 
corresponding EDA commands and evaluation queries. Each candidate is 
evaluated by saving a design checkpoint, applying the parameter modifications, executing the evaluation commands, and extracting the 
resulting metrics. 
The design is then restored to the checkpoint 
before the next candidate is evaluated.

The fitness of each candidate is computed as a stage-aware weighted 
score: $\text{score}(c) = \sum\nolimits_{m \in \mathcal{M}_s} w_m^{(s)} \cdot 
    \hat{v}_m(c)$,
where $\mathcal{M}_s$ denotes the metric set for stage $s$, 
$w_m^{(s)}$ the stage-specific weight, and $\hat{v}_m(c) \in [0, 1]$ 
the normalized metric value. The best candidate $c^* = 
\arg\max_c \text{score}(c)$ becomes the starting point for batch 
$b{+}1$. The search terminates when stage requirements are satisfied 
(e.g., DRC count $= 0$ for routing), the score improvement falls 
below threshold $\delta$, or the batch budget $B$ is exhausted.

\iffalse

\begin{equation}
    \text{score}(c) = \sum\nolimits_{m \in \mathcal{M}_s} w_m^{(s)} \cdot 
    \hat{v}_m(c),
\end{equation}

\fi

\noindent\textbf{Iteration Tracking.}
A persistent \texttt{MetricsDB} records the full parameter to metric 
mapping across all iterations, including parameter provenance, 
capturing which agent modified which parameter and the rationale 
behind each change. This trajectory data serves two purposes: it 
provides the Optimizer Agent with historical context for informed 
tuning decisions, and it enables post-hoc analysis of the DSO process 
for reproducibility and insight.

\section{Experiments}

% This section presents the experimental setup and evaluation results. 
We first describe the setup, including models, EDA tools, technology 
nodes, and server configuration. We then report the evaluation results, 
followed by a comparison against RAG-based method to further validate our proposed method.

\subsection{Experimental Setup}

\noindent\textbf{Models.} We evaluate both open-source and commercial 
LLMs. For open-source models, we select the multimodal model Qwen3-VL~\cite{qwen3-vlm} 
in 8B and 30B parameter variants. For commercial models, we use GPT-4o~\cite{gpt4o}, 
DeepSeek-V3~\cite{deepseek}, Gemini-3.1-pro-preview~\cite{gemini-3-pro}, and GPT-5-Thinking~\cite{gpt5}, accessed via 
their respective APIs. All models are evaluated at a temperature of 
0.1 unless otherwise specified.
\noindent\textbf{EDA Tools.} \textbf{The primary goal of our work is to benchmark the capabilities of LLMs/VLMs for VLSI physical design, not to compare or rank different EDA tools}. We use Cadence Innovus 22, Synopsys 
IC Compiler II (ICC2) 2022, Synopsys PrimeTime (PT) 2022, and Synopsys 
StarRC 2022 as our commercial EDA tools for implementation and signoff. 
Command-level differences across tool versions are minimal; we validate 
that our generated scripts are compatible with versions 2018 and 2020. 
\noindent\textbf{Technology Nodes.} For experiments using commercial 
EDA tools, we target the TSMC 28nm. 
\noindent\textbf{Platform.} All local models are deployed on a server equipped with 12 NVIDIA A6000 GPUs and an AMD EPYC 7313 CPU.

\begin{table}[t]
\centering
\caption{Evaluation on \textbf{OpenLayout-Bench} atomic task suite across 2 commercial place-and-route tools, anonymized as \emph{Tool~A} and \emph{Tool~B} (in arbitrary order), and one static timing analysis (STA) tool. Pass@1 and Pass@5 denote
the success rate on the first and fifth attempt, respectively.
The best result in each column is \textbf{bolded}.``--'' indicates that results are unavailable due to the absence of
multimodal input support in the corresponding model.
Gemini and DS refer to Gemini~3.1-pro-preview and DeepSeek-V3,
respectively.}
\label{tab:workflow_bench}
\vskip -9pt
\scalebox{0.88}{
\begin{tabular}{lcccccc}
\toprule
\multirow{2}{*}{Model} & \multicolumn{2}{c}{Tool~A} &
\multicolumn{2}{c}{Tool~B} & \multicolumn{2}{c}{STA} \\
\cmidrule(lr){2-3} \cmidrule(lr){4-5} \cmidrule(lr){6-7}
& P@1 & P@5 & P@1 & P@5 & P@1 & P@5 \\ 
\midrule
% \multicolumn{7}{l}{\textit{Standalone LLMs}} \\
% \midrule
Qwen3-VL-8B             & 0.122  & 0.156  & 0.136  & 0.182  & 0.182  & 0.364  \\
Qwen3-VL-30B            & 0.089  & 0.156  & 0.102  & 0.205  & 0.273  & 0.364  \\
DeepSeek-V3             & 0.167  & 0.200  & 0.182  & 0.227  & --     & --     \\
GPT-4o                  & 0.167  & 0.211  & 0.205  & 0.239  & 0.364  & 0.454  \\
Gemini-3.1-pro     & 0.189  & 0.222  & 0.193  & 0.216  & 0.636  & 0.727  \\
GPT-5-Thinking          & 0.178  & 0.222  & 0.205  & 0.261  & 0.545  & 0.636  \\
% \midrule
% \multicolumn{7}{l}{\textit{RAG-Enhanced LLMs (Ablation)}} \\
% \midrule
RAG-DS                  & 0.556  & 0.622  & 0.591  & 0.614  & --     & --     \\
RAG-Gemini              & 0.600  & 0.656  & 0.625  & 0.659  & 0.636  & 0.727  \\
% \midrule
% \multicolumn{7}{l}{\textit{OpenLayout (Ours)}} \\
% \midrule
OpenLayout-DS           & 0.778  & 0.822  & 0.807  & 0.841  & --     & --     \\
OpenLayout-Gemini       & \textbf{0.800}  & \textbf{0.844}  & \textbf{0.841}  & \textbf{0.864}  & \textbf{0.636}  & \textbf{0.727}  \\
OpenLayout-GPT5         & 0.789  & 0.822  & 0.818  & 0.852  & 0.545  & 0.636  \\
\bottomrule
\end{tabular} }
\vskip -12pt
\end{table}

\begin{table*}[t]
\centering
\caption{Full-flow implementation of \textbf{OpenLayout-Gemini} on 
the holistic flow suite (TSMC 28nm). WNS and DRC violations are reported 
at each stage; positive WNS indicates timing closure. Utilization (Util.)
is reported after floorplanning. The last column reports the total (input\,+\,output)
LLM token consumption of the entire flow.}
\label{tab:fullflow_results}
\vskip -9pt
\scalebox{0.8}{
% \usepackage{multirow}
\begin{tabular}{lllllllllc}
\hline
Design                               & init                                                                                              & floorplan                                                                        & powerplan                   & place                                                                                                  & CTS                                                                                                    & CTSopt                                                                                                & Route                                                                                                   & Routeopt                                                                                               & \begin{tabular}[c]{@{}c@{}}Total\\ tokens\end{tabular} \\ \hline
\multirow{3}{*}{TinyRISCV}           & \multirow{3}{*}{\begin{tabular}[c]{@{}l@{}}5ns, WNS: 1.362ns\\ Setup successfully\end{tabular}}   & \multirow{3}{*}{\begin{tabular}[c]{@{}l@{}}Util: 0.65\\ DRC Vio: 0\end{tabular}} & \multirow{3}{*}{DRC Vio: 0} & \multirow{3}{*}{\begin{tabular}[c]{@{}l@{}}WNS : -0.014ns\\ DRC Vio: 0\\ Density: 0.715\end{tabular}}  & \multirow{3}{*}{\begin{tabular}[c]{@{}l@{}}WNS : -0.127ns\\ DRC Vio: 0\\ Density: 0.7165\end{tabular}} & \multirow{3}{*}{\begin{tabular}[c]{@{}l@{}}WNS : 0.001ns\\ DRC Vio: 0\\ Density: 0.713\end{tabular}}  & \multirow{3}{*}{\begin{tabular}[c]{@{}l@{}}WNS : 0.021ns\\ DRC Vio: 1\\ Density: 0.722\end{tabular}}    & \multirow{3}{*}{\begin{tabular}[c]{@{}l@{}}WNS : 0.021ns\\ DRC Vio: 0\\ Density: 0.722\end{tabular}} & \multirow{3}{*}{334K} \\
                                     &                                                                                                   &                                                                                  &                             &                                                                                                        &                                                                                                        &                                                                                                       &                                                                                                         &                                                                                                        \\
                                     &                                                                                                   &                                                                                  &                             &                                                                                                        &                                                                                                        &                                                                                                       &                                                                                                         &                                                                                                        \\
\multirow{3}{*}{Open910}             & \multirow{3}{*}{\begin{tabular}[c]{@{}l@{}}1.5ns, WNS: 0.465ns\\ Setup successfully\end{tabular}} & \multirow{3}{*}{\begin{tabular}[c]{@{}l@{}}Util: 0.60\\ DRC Vio: 0\end{tabular}} & \multirow{3}{*}{DRC Vio: 0} & \multirow{3}{*}{\begin{tabular}[c]{@{}l@{}}WNS : -0.05ns\\ DRC Vio: 55\\ Density: 0.584\end{tabular}}  & \multirow{3}{*}{\begin{tabular}[c]{@{}l@{}}WNS : -0.009ns\\ DRC Vio: 2\\ Density: 0.653\end{tabular}}  & \multirow{3}{*}{\begin{tabular}[c]{@{}l@{}}WNS : 0.001ns\\ DRC Vio: 0\\ Density: 0.654\end{tabular}}  & \multirow{3}{*}{\begin{tabular}[c]{@{}l@{}}WNS : -0.431ns\\ DRC Vio: 127\\ Density: 0.703\end{tabular}} & \multirow{3}{*}{\begin{tabular}[c]{@{}l@{}}WNS : 0.008ns\\ DRC Vio: 125\\ Density: 0.731\end{tabular}} & \multirow{3}{*}{1.31M} \\
                                     &                                                                                                   &                                                                                  &                             &                                                                                                        &                                                                                                        &                                                                                                       &                                                                                                         &                                                                                                        \\
                                     &                                                                                                   &                                                                                  &                             &                                                                                                        &                                                                                                        &                                                                                                       &                                                                                                         &                                                                                                        \\
\multirow{3}{*}{Systolic array}      & \multirow{3}{*}{\begin{tabular}[c]{@{}l@{}}3.4ns, WNS: 1.11ns\\ Setup successfully\end{tabular}}  & \multirow{3}{*}{\begin{tabular}[c]{@{}l@{}}Util: 0.5\\ DRC Vio: 0\end{tabular}}  & \multirow{3}{*}{DRC Vio: 0} & \multirow{3}{*}{\begin{tabular}[c]{@{}l@{}}WNS : 0.013ns\\ DRC Vio: 0\\ Density: 0.464\end{tabular}}   & \multirow{3}{*}{\begin{tabular}[c]{@{}l@{}}WNS : 0.003ns\\ DRC Vio: 0\\ Density: 0.464\end{tabular}}   & \multirow{3}{*}{\begin{tabular}[c]{@{}l@{}}WNS : 0.003ns\\ DRC Vio: 0\\ Density: 0.464\end{tabular}}  & \multirow{3}{*}{\begin{tabular}[c]{@{}l@{}}WNS : -0.312ns\\ DRC Vio: 0\\ Density: 0.464\end{tabular}}   & \multirow{3}{*}{\begin{tabular}[c]{@{}l@{}}WNS : 0.002ns\\ DRC Vio: 0\\ Density: 0.466\end{tabular}} & \multirow{3}{*}{232K} \\
                                     &                                                                                                   &                                                                                  &                             &                                                                                                        &                                                                                                        &                                                                                                       &                                                                                                         &                                                                                                        \\
                                     &                                                                                                   &                                                                                  &                             &                                                                                                        &                                                                                                        &                                                                                                       &                                                                                                         &                                                                                                        \\
\multirow{3}{*}{16b pipelined mult.} & \multirow{3}{*}{\begin{tabular}[c]{@{}l@{}}2.6ns, WNS: 0.808ns\\ Setup successfully\end{tabular}} & \multirow{3}{*}{\begin{tabular}[c]{@{}l@{}}Util: 0.55\\ DRC Vio: 0\end{tabular}} & \multirow{3}{*}{DRC Vio: 0} & \multirow{3}{*}{\begin{tabular}[c]{@{}l@{}}WNS : 0.02ns\\ DRC Vio: 0\\ Density: 0.625\end{tabular}}    & \multirow{3}{*}{\begin{tabular}[c]{@{}l@{}}WNS : -0.008ns\\ DRC Vio: 0\\ Density: 0.6289\end{tabular}} & \multirow{3}{*}{\begin{tabular}[c]{@{}l@{}}WNS : 0.012ns\\ DRC Vio: 0\\ Density: 0.629\end{tabular}}  & \multirow{3}{*}{\begin{tabular}[c]{@{}l@{}}WNS : 0.251ns\\ DRC Vio: 0\\ Density: 0.629\end{tabular}}    & \multirow{3}{*}{\begin{tabular}[c]{@{}l@{}}WNS : 0.251ns\\ DRC Vio: 0\\ Density: 0.629\end{tabular}} & \multirow{3}{*}{148K} \\
                                     &                                                                                                   &                                                                                  &                             &                                                                                                        &                                                                                                        &                                                                                                       &                                                                                                         &                                                                                                        \\
                                     &                                                                                                   &                                                                                  &                             &                                                                                                        &                                                                                                        &                                                                                                       &                                                                                                         &                                                                                                        \\
\multirow{3}{*}{AES-256}             & \multirow{3}{*}{\begin{tabular}[c]{@{}l@{}}3.5ns, WNS: 1.003ns\\ Setup successfully\end{tabular}} & \multirow{3}{*}{\begin{tabular}[c]{@{}l@{}}Util: 0.60\\ DRC Vio: 0\end{tabular}} & \multirow{3}{*}{DRC Vio: 0} & \multirow{3}{*}{\begin{tabular}[c]{@{}l@{}}WNS : 0.003ns\\ DRC Vio: 0\\ Density: 0.613\end{tabular}}   & \multirow{3}{*}{\begin{tabular}[c]{@{}l@{}}WNS : 0.006ns\\ DRC Vio: 0\\ Density: 0.615\end{tabular}}   & \multirow{3}{*}{\begin{tabular}[c]{@{}l@{}}WNS : 0.006ns\\ DRC Vio: 0\\ Density: 0.615\end{tabular}}  & \multirow{3}{*}{\begin{tabular}[c]{@{}l@{}}WNS : 0.12ns\\ DRC Vio: 303\\ Density: 0.615\end{tabular}}   & \multirow{3}{*}{\begin{tabular}[c]{@{}l@{}}WNS : 0.119ns\\ DRC Vio: 0\\ Density: 0.615\end{tabular}} & \multirow{3}{*}{264K} \\
                                     &                                                                                                   &                                                                                  &                             &                                                                                                        &                                                                                                        &                                                                                                       &                                                                                                         &                                                                                                        \\
                                     &                                                                                                   &                                                                                  &                             &                                                                                                        &                                                                                                        &                                                                                                       &                                                                                                         &                                                                                                        \\
\multirow{3}{*}{NVDLA-small}         & \multirow{3}{*}{\begin{tabular}[c]{@{}l@{}}2ns, WNS: 0.65ns\\ Setup successfully\end{tabular}}    & \multirow{3}{*}{\begin{tabular}[c]{@{}l@{}}Util: 0.60\\ DRC Vio: 0\end{tabular}} & \multirow{3}{*}{DRC Vio: 0} & \multirow{3}{*}{\begin{tabular}[c]{@{}l@{}}WNS : -1.06ns\\ DRC Vio: 46\\ Density: 0.424\end{tabular}}  & \multirow{3}{*}{\begin{tabular}[c]{@{}l@{}}WNS : 0.002ns\\ DRC Vio: 0\\ Density: 0.554\end{tabular}}   & \multirow{3}{*}{\begin{tabular}[c]{@{}l@{}}WNS : 0.002ns\\ DRC Vio: 0\\ Density: 0.554\end{tabular}}  & \multirow{3}{*}{\begin{tabular}[c]{@{}l@{}}WNS : -0.005ns\\ DRC Vio: 21\\ Density: 0.556\end{tabular}}  & \multirow{3}{*}{\begin{tabular}[c]{@{}l@{}}WNS : 0.002ns\\ DRC Vio: 19\\ Density: 0.556\end{tabular}} & \multirow{3}{*}{1.28M} \\
                                     &                                                                                                   &                                                                                  &                             &                                                                                                        &                                                                                                        &                                                                                                       &                                                                                                         &                                                                                                        \\
                                     &                                                                                                   &                                                                                  &                             &                                                                                                        &                                                                                                        &                                                                                                       &                                                                                                         &                                                                                                        \\
\multirow{3}{*}{NVDLA-large}         & \multirow{3}{*}{\begin{tabular}[c]{@{}l@{}}2.5ns, WNS: 0.69ns\\ Setup successfully\end{tabular}}  & \multirow{3}{*}{\begin{tabular}[c]{@{}l@{}}Util: 0.60\\ DRC Vio: 0\end{tabular}} & \multirow{3}{*}{DRC Vio: 0} & \multirow{3}{*}{\begin{tabular}[c]{@{}l@{}}WNS : -0.024ns\\ DRC Vio: 68\\ Density: 0.487\end{tabular}} & \multirow{3}{*}{\begin{tabular}[c]{@{}l@{}}WNS : -0.145ns\\ DRC Vio: 0\\ Density: 0.586\end{tabular}}  & \multirow{3}{*}{\begin{tabular}[c]{@{}l@{}}WNS : -0.063ns\\ DRC Vio: 0\\ Density: 0.594\end{tabular}} & \multirow{3}{*}{\begin{tabular}[c]{@{}l@{}}WNS : 0.021ns\\ DRC Vio: 242\\ Density: 0.596\end{tabular}}  & \multirow{3}{*}{\begin{tabular}[c]{@{}l@{}}WNS : 0.021ns\\ DRC Vio: 85\\ Density: 0.598\end{tabular}} & \multirow{3}{*}{1.62M} \\
                                     &                                                                                                   &                                                                                  &                             &                                                                                                        &                                                                                                        &                                                                                                       &                                                                                                         &                                                                                                        \\
                                     &                                                                                                   &                                                                                  &                             &                                                                                                        &                                                                                                        &                                                                                                       &                                                                                                         &                                                                                                        \\
\multirow{3}{*}{UART controller}     & \multirow{3}{*}{\begin{tabular}[c]{@{}l@{}}3.4ns, WNS: 1.108ns\\ Setup successfully\end{tabular}} & \multirow{3}{*}{\begin{tabular}[c]{@{}l@{}}Util: 0.65\\ DRC Vio: 0\end{tabular}} & \multirow{3}{*}{DRC Vio: 0} & \multirow{3}{*}{\begin{tabular}[c]{@{}l@{}}WNS : 1.1ns\\ DRC Vio: 0\\ Density: 0.782\end{tabular}}     & \multirow{3}{*}{\begin{tabular}[c]{@{}l@{}}WNS : 1.4ns\\ DRC Vio: 0\\ Density: 0.789\end{tabular}}     & \multirow{3}{*}{\begin{tabular}[c]{@{}l@{}}WNS : 1.59ns\\ DRC Vio: 0\\ Density: 0.789\end{tabular}}   & \multirow{3}{*}{\begin{tabular}[c]{@{}l@{}}WNS : -0.445ns\\ DRC Vio: 0\\ Density: 0.85\end{tabular}}    & \multirow{3}{*}{\begin{tabular}[c]{@{}l@{}}WNS : -0.09ns\\ DRC Vio: 0\\ Density: 0.85\end{tabular}} & \multirow{3}{*}{132K} \\
                                     &                                                                                                   &                                                                                  &                             &                                                                                                        &                                                                                                        &                                                                                                       &                                                                                                         &                                                                                                        \\
                                     &                                                                                                   &                                                                                  &                             &                                                                                                        &                                                                                                        &                                                                                                       &                                                                                                         &                                                                                                        \\
\multirow{3}{*}{Elevator FSM}        & \multirow{3}{*}{\begin{tabular}[c]{@{}l@{}}2.5ns, WNS: 0.708ns\\ Setup successfully\end{tabular}} & \multirow{3}{*}{\begin{tabular}[c]{@{}l@{}}Util: 0.7\\ DRC Vio: 0\end{tabular}}  & \multirow{3}{*}{DRC Vio: 0} & \multirow{3}{*}{\begin{tabular}[c]{@{}l@{}}WNS : 0.703ns\\ DRC Vio: 2\\ Density: 0.664\end{tabular}}   & \multirow{3}{*}{\begin{tabular}[c]{@{}l@{}}WNS : 0.702ns\\ DRC Vio: 2\\ Density: 0.665\end{tabular}}   & \multirow{3}{*}{\begin{tabular}[c]{@{}l@{}}WNS : 0.703ns\\ DRC Vio: 0\\ Density: 0.667\end{tabular}}  & \multirow{3}{*}{\begin{tabular}[c]{@{}l@{}}WNS : 0.622ns\\ DRC Vio: 0\\ Density: 0.667\end{tabular}}    & \multirow{3}{*}{\begin{tabular}[c]{@{}l@{}}WNS : 0.622ns\\ DRC Vio: 0\\ Density: 0.667\end{tabular}} & \multirow{3}{*}{179K} \\
                                     &                                                                                                   &                                                                                  &                             &                                                                                                        &                                                                                                        &                                                                                                       &                                                                                                         &                                                                                                        \\
                                     &                                                                                                   &                                                                                  &                             &                                                                                                        &                                                                                                        &                                                                                                       &                                                                                                         &                                                                                                        \\
\multirow{3}{*}{Ethernet MAC}        & \multirow{3}{*}{\begin{tabular}[c]{@{}l@{}}10ns, WNS: 5.621ns\\ Setup successfully\end{tabular}}  & \multirow{3}{*}{\begin{tabular}[c]{@{}l@{}}Util: 0.60\\ DRC Vio: 0\end{tabular}} & \multirow{3}{*}{DRC Vio: 0} & \multirow{3}{*}{\begin{tabular}[c]{@{}l@{}}WNS : 5.62ns\\ DRC Vio: 25\\ Density: 0.62\end{tabular}}    & \multirow{3}{*}{\begin{tabular}[c]{@{}l@{}}WNS : 5.475ns\\ DRC Vio: 23\\ Density: 0.624\end{tabular}}  & \multirow{3}{*}{\begin{tabular}[c]{@{}l@{}}WNS : 5.470ns\\ DRC Vio: 0\\ Density: 0.628\end{tabular}}  & \multirow{3}{*}{\begin{tabular}[c]{@{}l@{}}WNS : 5.53ns\\ DRC Vio: 69\\ Density: 0.624\end{tabular}}    & \multirow{3}{*}{\begin{tabular}[c]{@{}l@{}}WNS : 5.53ns\\ DRC Vio: 0\\ Density: 0.625\end{tabular}} & \multirow{3}{*}{453K} \\
                                     &                                                                                                   &                                                                                  &                             &                                                                                                        &                                                                                                        &                                                                                                       &                                                                                                         &                                                                                                        \\
                                     &                                                                                                   &                                                                                  &                             &                                                                                                        &                                                                                                        &                                                                                                       &                                                                                                         &                                                                                                        \\ \hline
\end{tabular}

}
\vskip -9pt
\end{table*}

\subsection{Evaluation on Atomic Task Suite}

Evaluation results on \textbf{OpenLayout-Bench} atomic task suite are summarized in Table~\ref{tab:workflow_bench}. 
Standalone LLMs 
perform poorly across all tools; for example, even the powerful 
Gemini~3.1-pro achieves only 18.9\% on the Tool A task set and 
19.3\% on the Tool B task set, comparable to the much smaller 
Qwen3-VL-8B. 
We attribute this to the scarcity of EDA-related data 
in public corpora, as EDA tools and scripts are often proprietary and 
not widely available online.
Upon investigating the failure cases, we 
find that many errors are minor in nature, such as syntax mistakes 
or incorrect parameter usage, suggesting that the models demonstrate 
basic domain awareness and generate structurally reasonable scripts, 
but lack the precise tool-level knowledge required for successful 
execution. This observation further underscores the value of targeted 
skill augmentation, as even small corrections can unlock substantially 
higher pass rates.

Interestingly, standalone LLMs perform significantly better on the STA 
task set, which relies heavily on mathematical reasoning, indicating 
that these models possess strong general mathematical capability but 
lack domain-specific EDA knowledge.
In contrast, \textbf{OpenLayout} substantially improves pass rates 
across all models and tools. 
For example, \textbf{OpenLayout} built on 
Gemini~3.1-pro achieves 84.4\% on the Tool A task set and 
86.4\% on the Tool B task set, representing a relative improvement of 
 $>4\times$ over the standalone baseline. 
These results 
demonstrate that for domain-knowledge-intensive tasks such as physical 
design, an agentic framework equipped with domain-specific skills is 
essential.

\subsection{Evaluation on Holistic Flow Suite}

Evaluation results of \textbf{OpenLayout} on the 
OpenLayout-Bench holistic flow suite are summarized in Table~\ref{tab:fullflow_results}, 
using Gemini 3.1-pro-preview as the backbone model. The results show that \textbf{OpenLayout} is capable of executing a holistic physical 
design flow end-to-end, performing self-iterative design space exploration, 
autonomously resolving DRC violations, and optimizing the design toward 
timing closure. The consistent performance across diverse designs 
demonstrates the generality of our method. On average, \textbf{OpenLayout} 
completes the full flow through 53 interactions with the EDA tools, 
reflecting the iterative nature of physical design optimization.

Our method performs well on small and medium-complexity designs, 
achieving timing closure with zero or minimal DRC violations at the 
final Routeopt stage. However, on larger and more complex designs such 
as NVDLA-large and Open910, residual DRC violations remain after 
Routeopt, and timing closure is not fully achieved in all cases. 
We attribute these limitations not only to the inherent complexity 
of the design space at advanced technology nodes, but also to the 
current lack of more sophisticated ECO and DRC fixing capabilities 
in our framework, which we identify as a promising direction for 
future work. We anticipate that future work building upon 
\textbf{OpenLayout} can achieve improved timing slack, higher 
utilization, cleaner DRC results, and reduced interaction overhead 
through more advanced optimization strategies.

\subsection{Comparison Against RAG Baseline}
\label{rag_baseline}

We further evaluate two RAG-enhanced LLM baselines, which retrieve 
relevant content from EDA command reference manuals and physical design 
guidebooks, using Gemini~3.1-pro-preview and DeepSeek-V3 as backbone 
models for fair comparison. Atomic task suite results are summarized in 
Table~\ref{tab:workflow_bench} (RAG-DS and RAG-Gemini rows), and 
holistic flow suite results are reported in 
Table~\ref{tab:rag_results}.

\begin{table}[]
\caption{Early-stage physical design results using a RAG-based LLM method 
on the OpenLayout-Bench circuit suite (TSMC 28nm). Results cover the first 
three implementation stages: synthesis, floorplanning, and power planning. 
The RAG-based approach exhibits consistent failures across all stages, 
including setup errors, pin placement failures, and incomplete 
power networks, demonstrating its limited capability in automated physical 
design.}
\label{tab:rag_results}
\vskip -9pt
\scalebox{0.80}{
\begin{tabular}{llll}
\hline
Design                               & init                                                                                              & floorplan                                                                        & powerplan                                                                                          \\ \hline
\multirow{3}{*}{TinyRISCV}           & \multirow{3}{*}{\begin{tabular}[c]{@{}l@{}}5ns, WNS: 1.362ns\\ Setup failed\end{tabular}}         & \multirow{3}{*}{\begin{tabular}[c]{@{}l@{}}Pin placement \\ failed\end{tabular}} & \multirow{3}{*}{\begin{tabular}[c]{@{}l@{}}Failed to use \\ the correct \\ layer\end{tabular}}     \\
                                     &                                                                                                   &                                                                                  &                                                                                                    \\
                                     &                                                                                                   &                                                                                  &                                                                                                    \\
\multirow{3}{*}{Open910}             & \multirow{3}{*}{\begin{tabular}[c]{@{}l@{}}1.5ns, WNS: 0.465ns\\ Setup failed\end{tabular}}       & \multirow{3}{*}{\begin{tabular}[c]{@{}l@{}}Pin placement \\ failed\end{tabular}} & \multirow{3}{*}{\begin{tabular}[c]{@{}l@{}}Failed to \\ addRing\end{tabular}}                      \\
                                     &                                                                                                   &                                                                                  &                                                                                                    \\
                                     &                                                                                                   &                                                                                  &                                                                                                    \\
\multirow{3}{*}{Systolic array}      & \multirow{3}{*}{\begin{tabular}[c]{@{}l@{}}3.4ns, WNS: 1.11ns\\ Setup successfully\end{tabular}}  & \multirow{3}{*}{\begin{tabular}[c]{@{}l@{}}Pin placement \\ failed\end{tabular}} & \multirow{3}{*}{DRC Vio: 346}                                                                      \\
                                     &                                                                                                   &                                                                                  &                                                                                                    \\
                                     &                                                                                                   &                                                                                  &                                                                                                    \\
\multirow{3}{*}{16b pipelined mult.} & \multirow{3}{*}{\begin{tabular}[c]{@{}l@{}}2.6ns, WNS: 0.808ns\\ Setup failed\end{tabular}}       & \multirow{3}{*}{\begin{tabular}[c]{@{}l@{}}Util: 0.65\\ DRC Vio: 0\end{tabular}} & \multirow{3}{*}{\begin{tabular}[c]{@{}l@{}}Missing Bottom \\ and Top Power \\ Rings\end{tabular}}  \\
                                     &                                                                                                   &                                                                                  &                                                                                                    \\
                                     &                                                                                                   &                                                                                  &                                                                                                    \\
\multirow{3}{*}{AES-256}             & \multirow{3}{*}{\begin{tabular}[c]{@{}l@{}}3.5ns, WNS: 1.003ns\\ Setup failed\end{tabular}}       & \multirow{3}{*}{\begin{tabular}[c]{@{}l@{}}Pin placement \\ failed\end{tabular}} & \multirow{3}{*}{\begin{tabular}[c]{@{}l@{}}Stripes Not \\ Connected to \\ Power Ring\end{tabular}} \\
                                     &                                                                                                   &                                                                                  &                                                                                                    \\
                                     &                                                                                                   &                                                                                  &                                                                                                    \\
\multirow{3}{*}{NVDLA-small}         & \multirow{3}{*}{\begin{tabular}[c]{@{}l@{}}2ns, WNS: 0.65ns\\ Setup failed\end{tabular}}          & \multirow{3}{*}{\begin{tabular}[c]{@{}l@{}}Pin placement \\ failed\end{tabular}} & \multirow{3}{*}{\begin{tabular}[c]{@{}l@{}}Missing Bottom \\ and Top Power \\ Rings\end{tabular}}  \\
                                     &                                                                                                   &                                                                                  &                                                                                                    \\
                                     &                                                                                                   &                                                                                  &                                                                                                    \\
\multirow{3}{*}{NVDLA-large}         & \multirow{3}{*}{\begin{tabular}[c]{@{}l@{}}2.5ns, WNS: 0.69ns\\ Setup failed\end{tabular}}        & \multirow{3}{*}{\begin{tabular}[c]{@{}l@{}}Pin placement \\ failed\end{tabular}} & \multirow{3}{*}{\begin{tabular}[c]{@{}l@{}}Missing Bottom \\ and Top Power \\ Rings\end{tabular}}  \\
                                     &                                                                                                   &                                                                                  &                                                                                                    \\
                                     &                                                                                                   &                                                                                  &                                                                                                    \\
\multirow{3}{*}{UART controller}     & \multirow{3}{*}{\begin{tabular}[c]{@{}l@{}}3.4ns, WNS: 1.108ns\\ Setup successfully\end{tabular}} & \multirow{3}{*}{\begin{tabular}[c]{@{}l@{}}Util: 0.65\\ DRC Vio: 0\end{tabular}} & \multirow{3}{*}{\begin{tabular}[c]{@{}l@{}}Stripes Not \\ Connected to \\ Power Ring\end{tabular}} \\
                                     &                                                                                                   &                                                                                  &                                                                                                    \\
                                     &                                                                                                   &                                                                                  &                                                                                                    \\
\multirow{3}{*}{Elevator FSM}        & \multirow{3}{*}{\begin{tabular}[c]{@{}l@{}}2.5ns, WNS: 0.708ns\\ Setup successfully\end{tabular}} & \multirow{3}{*}{\begin{tabular}[c]{@{}l@{}}Util: 0.7\\ DRC Vio: 0\end{tabular}}  & \multirow{3}{*}{\begin{tabular}[c]{@{}l@{}}Stripes Not \\ Connected to \\ Power Ring\end{tabular}} \\
                                     &                                                                                                   &                                                                                  &                                                                                                    \\
                                     &                                                                                                   &                                                                                  &                                                                                                    \\
\multirow{3}{*}{Ethernet MAC}        & \multirow{3}{*}{\begin{tabular}[c]{@{}l@{}}10ns, WNS: 5.621ns\\ Setup failed\end{tabular}}        & \multirow{3}{*}{\begin{tabular}[c]{@{}l@{}}Pin placement \\ failed\end{tabular}} & \multirow{3}{*}{\begin{tabular}[c]{@{}l@{}}Missing Bottom \\ and Top Power \\ Rings\end{tabular}}  \\
                                     &                                                                                                   &                                                                                  &                                                                                                    \\
                                     &                                                                                                   &                                                                                  &                                                                                                    \\ \hline
\end{tabular} }
\vskip -12pt
\end{table}

Although RAG-enhanced methods improve over standalone LLMs, 
RAG-Gemini achieves 60.0\% Pass@1 on Innovus compared to 18.9\% 
for standalone Gemini, they still fall substantially short of 
\textbf{OpenLayout}, which achieves 80.0\% Pass@1 on the same tool. 
For the holistic flow suite, the RAG-enhanced method fails to complete 
most designs beyond the first two stages. We attribute this 
performance gap to the highly structured and interdependent nature 
of physical design flows: each stage comprises multiple steps with 
precise tool-level syntax and context-specific requirements, which 
general-purpose retrieved documentation is too coarse-grained to 
fully capture. Unlike \textbf{OpenLayout}, which encodes actionable, 
stage-specific skills, the RAG-based approach lacks the procedural 
knowledge necessary to handle tool-specific errors, and adapt parameters 
across designs, and drive the flow to completion.

\subsection{Limitations and Future Work}

\noindent\textbf{Limitations.} Our platform currently covers the 
physical design workflow from initialization to ECO, but has several 
notable limitations. First, LVS (Layout versus Schematic) verification 
is not yet supported, leaving an important signoff step unaddressed. 
Second, DRC fixing remains incomplete, particularly for violations 
that require GUI-based interactions, which are difficult to automate 
through scripted flows. Third, our evaluation is currently limited 
to the TSMC 28nm technology node, which may not fully reflect 
performance across other process nodes. Finally, the 
\textbf{OpenLayout-Bench} suite would benefit from a broader 
collection of large-scale industrial VLSI designs to more 
comprehensively stress-test the framework.
\noindent\textbf{Future Work.} We plan to extend \textbf{OpenLayout} 
along several directions. First, we will expand technology node 
support beyond 28nm to include both advanced nodes (e.g., 16nm, 7nm) 
and mature nodes (e.g., 65nm, 130nm), enabling broader applicability. 
Second, we will enrich \textbf{OpenLayout-Bench} with larger and more 
complex VLSI designs to improve benchmark coverage and difficulty. 
Third, we will develop more fine-grained, stage-specific optimization 
skills and algorithms, such as automated macro placement and congestion-driven 
floorplanning.

\section{Conclusion}

In this work, we introduce \textbf{OpenLayout}, an open agentic 
platform for automated physical design, along with 
\textbf{OpenLayout-Bench}, a comprehensive benchmark suite for 
evaluating LLM agent capabilities at advanced technology nodes. 
OpenLayout supports a holistic physical design flow from 
initialization to Routeopt, enabling efficient iterative design space 
exploration, autonomous violation fixing, and timing closure across 
diverse circuit designs. Extensive experiments demonstrate that 
OpenLayout achieves SOTA performance, 
substantially outperforming both standalone LLMs and RAG-enhanced 
baselines. We hope that our work will inspire further research into 
agentic frameworks for fully automated physical design, encouraging 
the community to contribute additional benchmarks and domain-specific 
knowledge to advance the field.
% This work is supported in part by NSF grants under OAC-2514071 and OAC-2613629.

\section{Acknowledgments}
This work is supported in part by NSF grants under OAC-2514071 and OAC-2613629.

\bibliographystyle{ACM-Reference-Format}
\bibliography{references}

\appendix

\end{document}